\documentclass[useAMS,usenatbib,fleqn]{mnras}
\usepackage{amsmath}
\usepackage{multirow}
\usepackage{graphicx}
\usepackage{lipsum}

\usepackage{color,soul}
\usepackage{epsfig}
\usepackage{amssymb,amsmath}
\usepackage{aas_macros}
\usepackage{fixltx2e}
\global\long\def\Mo{M_{\odot}}

\global\long\def\Mbh{M_{\rm bh}}
\global\long\def\Ms{M_{\star}}

\global\long\def\Ns{N_{\star}}

\global\long\def\ss{\sigma_{\star}}
\global\long\def\sbh{\sigma_{\rm bh}}
\global\long\def\dbh{\Delta_{\rm bh}}

\global\long\def\vbh{v_{\rm bh}}
\global\long\def\rbh{r_{\rm bh}}

\global\long\def\vrel{v_{\mathrm{rel}}}
\global\long\def\Medd{\dot{M}_{E}}
\global\long\def\Mdot{\dot{M}}
\global\long\def\rtr{r_{\rm trap}}
\global\long\def\dinf{\rho_{\infty}}
\global\long\def\cinf{c_{\infty}}
\global\long\def\tinf{t_{\infty}}

\global\long\def\Mbh{M_{\rm bh}}

\title[A new channel to form IMBHs throughout cosmic time]{
A new channel to form IMBHs throughout cosmic time}

\author[Natarajan]{Priyamvada Natarajan${^{1,2,3}}$ \\
$^1$ Department of Astronomy, Yale University,  New Haven, CT 06511 \\
$^2$ Department of Physics, Yale University, New Haven, CT 06511 \\
$^3$ Yale Center for Astronomy \& Astrophysics, Yale University, New Haven, CT 06520\\
}

\date{\today}
\begin{document}
\pagerange{\pageref{firstpage}--\pageref{lastpage}} \pubyear{2017}
\maketitle

\begin{abstract}
While the formation of the first black holes at high redshift is reasonably well understood though debated, massive black hole (BH) formation at later cosmic epochs has not been adequately explored. We present a gas accretion driven mechanism that can build up black hole BH masses rapidly in dense, gas-rich nuclear star clusters (NSCs). Wind-fed supra-exponential accretion in these environments under the assumption of net zero angular momentum for the gas, can lead to extremely rapid growth, scaling stellar mass remnant seed BHs up to the intermediate mass black hole (IMBH) range. This new long-lived channel for IMBH formation permits growth to final masses ranging from $50 - 10^5\Mo$. Growth is modulated by the gas supply, and premature termination can result in the formation of BHs with masses between 50 - a few 100$\Mo$ filling in the so-called mass gap. Typically, growth is unimpeded and will result in the formation of IMBHs with masses ranging from $\sim 100 - 10^5 \Mo$. New detections from the LIGO-VIRGO source GW190521 to the emerging population of $\sim 10^5\Mo$ BHs harbored in low-mass dwarf galaxies - are revealing this elusive population.  Naturally accounting for the presence of off-center BHs in low-mass dwarfs, this new pathway also predicts the existence of a population of wandering non-central BHs in more massive galaxies detectable via tidal disruption events and as GW coalescences. Gas rich NSCs could therefore serve as incubators for the continual formation of BHs over a wide range in mass throughout cosmic time.
\end{abstract}

\begin{keywords}
early universe---black hole physics---accretion---Stars: dynamics
and kinematics---Galaxy: formation---galaxies: high-redshift
\end{keywords}

\section{Introduction}

\label{sec:introduction}

Our universe appears to be littered with black holes, near and far, active and dormant, with masses ranging from a few $\Mo$ to supermassive ones weighing $10^{6-10}\,\Mo$. Bridging these mass ranges, an expected population of intermediate mass black holes (IMBHs) with masses in the range $100-10^5\,\Mo$ are predicted to exist \citep{Rees1984,Bond+1984,Madau+2001}. This population of IMBHs has been challenging to detect, but they are now slowly starting to be uncovered. The more massive IMBHs with masses of $\sim 10^5\Mo$ hosted in low-mass and low-luminosity galaxies have been detected across a range of wavelengths from X-ray to radio \citep{Reines12,Baldassare15,Lemons15, Pardo16,Nucita+2017,Nucita+2018,Chilingarian+2018,Baldassare18,Reines+20,Mezcua+20}. In addition, the recent detection of the LIGO-VIRGO source GW190521 - a merger of two black holes with masses $\sim$ 85 $\Mo$ and 66 $\Mo$ and a final merged mass of $\sim$ 142 $\Mo$  - has revealed the existence of black holes at the low mass end of the IMBH mass range (\cite{LIGO+2020}). Intriguingly, the mass of the primary in GW190521 lies in the so-called pair instability supernova mass gap, the mass range between  50 - 150 $\Mo$ in which stellar processes are not expected to leave behind any compact remnants \cite{Woosley2017}. The large mass of the primary black hole in the GW190521 merger puts it right in the mass gap and therefore necessitates alternate formation mechanisms. 

Observational detection of accreting black holes has primarily been in electromagnetic wavelengths, while more recently detection via gravitational wave (GW) emission during BH-BH coalescences by the LIGO-VIRGO collaboration has opened up a new window into the mass function of stellar mass black holes and potentially IMBHs \citep{LIGO+2016,LIGO+2020}. Accreting supermassive black holes (SMBHs) have been typically detected as quasars and AGN at high and moderate redshifts, while accreting stellar mass black holes, are typically detected in the nearby universe as X-ray binaries. In our own Galaxy, a total of ${10^6-10^8}$ stellar mass black holes are predicted to exist, however, only a hundred or so X-ray binaries are seen, therefore it is anticipated that a vast population of dormant stellar mass black holes might be lurking (see review by \cite{RemillardMcClintock2006}). Obtaining a full census of non-accreting stellar mass black holes has been challenging, though the recent detection of unusual triple systems like HR 6819 are likely revealing a new repository of dormant stellar mass remnants \citep{Rivinius+20}. The prospects for a more complete multi-messenger census of stellar mass black holes looks extremely promising.

While for dormant SMBHs indirect detection via their dynamical influence on stars orbiting in their vicinity in galactic centers is viable for nearby galaxies, we need to rely on occasional luminous tidal disruption events (TDEs) - stripping of stars that stray close - for more distant galaxies \citep{Rees1988}. Once again, with advances in time-domain astronomy - more sophisticated survey strategies and dedicated instrumentation - there has been tremendous progress in the detection of transient tidal disruption flares. At the present time, several dozen TDEs have been observed with multi-wavelength surveys and analyzed (c.f. recent reviews by \cite{Komossa15,stone+17}). 

The local demography of SMBHs suggests that most galaxies in the universe harbor them at their centers \citep{Kormendy&Richstone1995}. SMBHs are detected in galactic centers using a multiplicity of observational techniques using gas, stellar, or maser dynamics as tracers \citep{Saglia+2016,vandenBosch2016}, including Sgr A* at the centre of our own galaxy \citep{Ghez+2008,Genzel+2010,Gillessen+2016}. Studies have also revealed correlations between BH mass and various host galaxy properties, like the stellar mass and luminosity of the bulge, however, the tightest correlation is the so-called  $M_\bullet-\sigma$ relation, that relates the mass of the central SMBH to the velocity dispersion of stars in inner-most regions of the host galaxy \citep{Ferrarese02,Tremaine+2002,Kormendy&Ho2013}. These relations, it was surmised, are likely the result of SMBH and host galaxy coevolution in the context of the currently accepted LCDM model. While the idea of co-evolution has increasingly garnered support, it remains unknown how and precisely when, in terms of cosmic epoch, these relations are established; how far down in black hole mass they extend to and if they evolve with redshift. The origin of these correlations, and whether they belie underlying physical causation mechanisms, perhaps an imprint of the initial conditions that reflect the formation of initial black hole seeds or their growth history over cosmic time is currently debated (see review for discussion of these open questions and references therein \cite{Natarajan2014}). How far down in host galaxy mass and hence central black hole mass, these local relations extend to is also unsettled at present as it gets progressively more challenging to detect accreting lower mass black holes - IMBHs - in nearby low mass galaxies as they are exceedingly faint. In the context of the larger structure formation scenario in a LCDM universe, within which the black hole growth paradigm is embedded into galaxy assembly, in the hierarchical, merger driven model, one set of explanations originally proposed for the observed correlation is that SMBHs and their hosts grow in concert over cosmic time (as suggested bt \cite{Haehnelt+1998,Kauffmann&Haehnelt2000,Volonteri+2003,Somerville+2008} and many other authors subsequently). This, it is argued, happens when a major merger between galaxies funnels gas to the centre of the galaxy, that both triggers an active galactic nucleus (AGN) and a burst of star formation \citep{DiMatteo+2005}.  One empirical clue supporting this picture is the observed ``AGN main sequence'' a linear correlation between SMBH accretion rates and star formation rates inferred from fitting spectral energy distributions, detected by stacking star-forming galaxies \citep{Mullaney+2012}. 

It has been challenging observationally to unravel the link between galaxy mergers and AGN activity, see for example \citep{Mechtley+2016} and recent review by \cite{AlexanderRev+2012}.  It has been proposed that perhaps it is only the most luminous AGN that are triggered via major mergers \citep{Treister+2012,Hong+2015,Marian+2019}, although this matter is far from settled \citep{Villforth+2017}. Some authors have suggested that these empirical correlations seen locally could simply be a consequence of the central limit theorem: a correlation can result simply from many mergers of initially uncorrelated SMBHs and host galaxies that occur over cosmic time \citep{Peng2007,Jahnke&Maccio2011}. 

Whether this correlation extends with the same observationally determined slope all the way down to IMBH masses or whether it flattens at some characteristic mass scale, is yet to be determined \citep{Greene2012}. Till recently, the issue of whether low surface brightness, low mass galaxies harbor central black holes at all was unsettled. New observational studies have been steadily uncovering accreting IMBHs $\sim\,10^5\,\Mo$ black holes - faint AGN - in low mass, dwarf galaxies \citep{Reines12,Baldassare15,Lemons15, Pardo16,Nucita+2017,Nucita+2018,Baldassare18,Reines+20}. These AGN detections out to $z \sim 3$,  are hosted in either disky or dwarf galaxies with stellar masses $M_*\,\leq \, 10^9\,\Mo$ and have been detected via their narrow emission line diagnostics as well as broad lines that have enabled their mass estimates; X-ray and radio emission signatures as well as the presence of higher ionization emission lines and on the basis of optical variability (\cite{Mezcua+2018,Mezcua+20}). In several instances, as shown in this most recent compilation, \cite{Reines+20} find that these accreting sources lie off-center, displaced from the galactic nucleus. 

In cosmological simulations tailored to track the growth history of black holes in dwarf galaxies, \cite{Bellovary+19} for instance, find that their occupation fraction depends on the host mass (massive dwarfs are more likely to host black holes); 50\% of the black holes are not centrally located and are wandering around within a few kpc of the center and given their low accretion luminosities are unlikely to be directly detected via emission. The GW signatures from their mergers though, would serve as the optimal way to detect the population, and this is the merger mass range at which the planned LISA mission is expected to have maximum sensitivity (\cite{LISA+2017,Bellovary+19}).

These growing discoveries of the long elusive IMBHs though raises questions about their formation. It is natural to assume that the initial black hole seeds formed at high redshifts whose growth was stunted causing them to not evolve to SMBHs at later times are the likely origin for IMBHs. Early modeling work suggested that the occupation fraction of these low mass black holes hosted in low mass, low luminosity hosts might therefore even offer clear glimpses of the pristine signatures of the massive seed formation channel \cite{Volonteri&Natarajan2009}. However, it has become clear in subsequent work, both in simulations and from observations that feedback from the accretion process might significantly alter black hole growth in these dwarf galaxies rendering these to no longer be untouched relics of high redshift seeds \citep{Mezcua19}. Though some signatures of seeding are predicted to persist, however, disentangling them from uncertainties in our knowledge of the accretion physics is complicated \citep{Ricarte&Natarajan2018a}. Meanwhile, the case for preferential stunted growth in low mass host haloes can be made, this though is applicable only to the small fraction that have experienced quieter merger histories (Natarajan et al. 2020, in prep.). It is entirely possible that IMBHs have remained elusive as they are preferentially located off-center or inhabit the outer regions of galaxies slowly grinding their way in as wanderers which might eventually merge or as lingerers that would take longer than a Hubble to make their way in. Another possibility that has been suggested is that the IMBH stage during the assembly of SMBHs might be short-lived due to optimal conditions that permit rapid growth \citep{Pacucci+2017}. Both off-center black hole populations - wanderers and lingerers - are currently detected in simulations like the Romulus Suite (Ricarte et al. 2020, in prep.). The emerging understanding of growing AGN in dwarf galaxies coupled with the fact that many of these accreting sources are found to be off-center, necessitates a new and alternate channel for their continual formation and growth down to late cosmic times \citep{Mezcua19,Mezcua+20}. 

In this paper, we propose a physical mechanism that could in principle, operate continually over cosmic time to produce a wide mass range of black holes from 50 - 150 $\Mo$ and all the way up to $10^{4-5}\,\Mo$ IMBHs by amplifying growth of initial stellar mass remnant black holes inside gas rich NSCs. We find that such an origin can account for detected IMBH masses; their off-center spatial locations and perhaps even account for the observed occurrence in special cases of central SMBHs embedded in NSCs as in the case of the Milky Way \citep{Schodel+2008}.

While observationally we are gradually accessing the full range of the black hole mass function from stellar to intermediate to supermassive to ultra-massive black hole masses, theoretically many critical questions remain open: the origin of the first seed black holes, in particular, those that are likely to be progenitors of SMBHs; and crucially whether seed formation ceases or continues to late cosmic times. There are now two accepted formation routes that produce light and massive initial black hole seeds at the earliest epochs in the universe. The primary formation mode for light seeds is believed to be from the remnants of the first generation of stars (Pop III stars) that likely formed from pristine gas with a top-heavy initial mass function (IMF), leaving behind black holes with masses ranging from a few $\Mo$ to about a 100 $\Mo$, depending on the details of the IMF \citep{Abel02,Bromm04,Hirano+2014}. Massive seeds in the IMBH mass range weighing $10^{4-5}\,\Mo$, on the other hand are proposed to have been produced from the direct collapse of pristine, pre-galactic gas disks or from super-Eddington accretion onto light seeds (see review by \cite{Natarajan2014} and references therein). Current models are firmly rooted in the early assembly and subsequent growth of these seeds and scant attention has been paid to whether continual IMBH formation might be permitted {\bf at later epochs}. Work by \cite{stone+17} explores this possibility of the formation of massive black holes in low redshift galactic nuclei, in NSCs from runaway tidal encounters. We summarize key aspects of this particular model as it is expected to proceed in the same sites with gas facilitated dynamics - in gas-rich NSCs - and can feasibly produce IMBHs continually over cosmic time akin to our proposed mechanism in which however, the mass assembly is propelled by gas accretion.

Building upon and extensively expanding on the model originally proposed by \cite{AlexanderNatarajan2014}, here we present a formation channel that can feasibly operate throughout cosmic time in gas-rich NSCs and would permit amplified growth by gas accretion for initially stellar mass black hole seeds. NSCs with the appropriate central densities and gas content are the perfect incubators for the growth of black holes via supra-exponential accretion, details that we present and explore further here. In \S 2, we summarize the current understanding of the formation of IMBHs in the very early universe followed by a brief review previous work on NSCs as sites of BH formation in \S 3, that includes a summary of observed properties of NSCs, and the key features of several models proposed by other workers for IMBH formation via merging processes and runaway tidal encounters. The physical mechanism proposed here for the continual formation of black holes over time by supra-exponential accretion in NSCs is described in detail in \S 5. Results predicted by this model are presented in \S 6. The conclusions and future prospects are outlined in the final section that includes a discussion of the implications of this new channel (i)  for the formation of the massive end of the mass function of stellar mass black holes and the IMBH recently detected by LIGO-VIRGO; (ii) for the recently detected IMBHs in low mass galaxies and (iii) the implications for the occupation fraction; prevalence of wandering as well as lingering black holes in galaxies and the expected enhancement in LISA merger rates.

\section{Summary of high redshift IMBH formation models}

There are currently two classes of seeding models that produce light and massive initial seeds at high redshift - both offer pathways for making IMBHs at early cosmic epochs.  A brief summary of these high redshift formation channels is warranted as one of them, the direct collapse model produces IMBHs from the get go. Meanwhile, light seeds following growth in early galaxies reach IMBH masses. Given these two formation channels, we expect IMBHs to be fairly abundant yet challenging to detect directly in the early universe. Light stellar mass seeds originate as the end state of the first generation Pop III stars.  The first episode of star formation in the early LCDM universe $(z \geq 15)$, commenced with the cooling and condensing of pristine gas in collapsed dark matter halos. At these epochs, atomic hydrogen was the only available coolant resulting in large Jeans masses for the collapsing gas causing the initial mass function of the first stars to be tilted high. This was observed in early computational models developed by \cite{Abel02,Bromm04}. However, in more recent simulations with higher resolution, gas clouds are seen to fragment more easily, leading to the formation of a cluster of lower mass stars \citep{Hirano+2014}. At any rate, as a consequence of stellar evolution, this first generation of stars is expected to leave behind remnant black holes - light initial seeds - with masses of a few to few tens of $\Mo$ \citep{Skinner+2020,Hirano+2014}. These light black hole seeds have been shown to subsequently grow too sub-optimally to account for the masses of the SMBHs powering observed quasars at $z  > 6$. This is due to the fact that their typical cosmic environments permit only sub-Eddington growth rates with short duty-cycles \citep{Alvarez+2009}. However, under some special circumstances, initially light stellar mass seeds can undergo brief periods of amplified growth at early cosmic epochs powered by super-Eddington accretion or grow via mergers with other light seeds as suggested by \cite{Volonteri&Rees2005} to account for the inferred masses powering the luminous quasar population.  Regardless, these light seeds  can and likely comfortably grow to the IMBH mass range, though direct detection via their accretion luminosities at these early epochs is extremely challenging with current observational facilities. Proposed future facilities like the Lynx Observatory will have the capability to detect IMBHs with masses $\leq 10^5 \Mo$ at $z \sim 10$ \footnote{For more details see Lynx Concept Study available at: https://wwwastro.msfc.nasa.gov/lynx/docs/LynxConceptStudy.pdf.}.

Massive seeds, whose birth masses ($10^{4-5}\,\Mo$) lie in the IMBH range, are postulated to originate from the direct collapse of low angular momentum primordial gas clouds as proposed by \citet{Koushiappas04,Eisenstein95,Begelman+2006,Lodato&Natarajan2006,Lodato&Natarajan2007}. The formation of these direct collapse black holes (DCBHs) requires the following special physical conditions in preferentially low angular momentum host halos: the presence of metal-free, pristine gas in these halos that settles into a massive disk that goes globally unstable via induced non-axisymmetric instabilities like bars; and the presence of Lyman-Werner photons that dissociate H$_2$ molecules to suppress cooling and fragmentation to prevent the formation of stars \citep{Oh&Haiman2002}. DCBHs form when gas in these massive, early pre-galactic disks goes dynamically unstable and rapidly drains matter to the center dispersing angular momentum outward via small scale bars and consequently growing a massive central object. In a low angular momentum setting, wherein disk fragmentation is prevented, the direct collapse of gas produces seeds with IMBH masses $\sim\,{10^{4-5}} \Mo$, significantly more massive than those generated by Pop III stars \citep[e.g.,][]{Clark+2008,Ferrara+2014,Agarwal+2016}. 

\cite{Lodato&Natarajan2006, Lodato&Natarajan2007} connected the large-scale cosmological context to the properties of these assembling seeds and derived a predicted the high redshift initial mass function for these IMBH seeds. The prediction of the birth mass function of these massive seed black holes has permitted demographic modeling and tracking their growth and assembly history over cosmic time. An interesting feature of these models is a natal correlation between properties of the host galactic nucleus and initial massive seed mass, i.e. the $M_{\rm bh}-\sigma$ correlation originates as a consequence of initial conditions (\cite{Volonteri+2008}). Current state-of-the-art simulations find that sites that satisfy the above criteria for pre-galactic direct collapse are available in the LCDM cosmogony, and have the appropriate abundance to account for the observed rare luminous quasars at high redshift (\cite{Agarwal+2014}). In fact, the massive seed formation channel was originally motivated in order to explain the bright, rare quasar populations detected at $z > 6$ from surveys like the Sloan Digital Sky Survey \citep{Fan03}. The more ubiquitous Pop III remnant light seeds however, serve as the progenitors for the typical SMBH with masses in the range of $10^{6-7}\,\Mo$ detected today. Therefore, with the head-start in initial masses, while the massive IMBH seeds can comfortably account for the SMBHs powering the detected luminous quasar population at the highest redshifts, while the light seeds suffice and can adequately explain the rest of the quasar/AGN population as well as the locally detected dormant SMBH population. An alternate pathway that produces massive initial seeds with masses in the IMBH range via the dynamical core collapse of the first generation of star clusters  at early times has been proposed by \cite{Lupi+2014}.

\section{Nuclear Star Clusters as sites of black hole formation}

NSCs and SMBHs are often detected at the centers of galaxies, typically in galaxies with $1{0^9\,M_{\odot}\, < M_* < 10^{11}\,M_{\odot}}$ and they can often co-exist (see \cite{georgiev+16} and references therein). However, in lower mass galaxies with $M_* < 10^{11}\,M_{\odot}$ galactic nuclei contain NSCs with masses that are proportional the stellar mass of the spheroidal component; and in the most massive galaxies (that tend to not contain NSCs) its the SMBH that inhabits the galactic nucleus. Given that NSCs are ubiquitous, they have been proposed as sites for black hole seed formation, particularly at high redshift. Below, we briefly outline two previously proposed scenarios in which the formation of an IMBH in the mass range of $10^{3-5}\,\Mo$ is facilitated by various stellar dynamical processes and accretion that can occur in NSCs \citep{stone+17,devecchi+09} and one scenario that implicates gas dynamical processes in NSCs proposed by \cite{Lupi+2014}. Recently \cite{Fragione+2020} have proposed a merger driven model operating in NSCs that yields a final massive BH with mass $\sim 400- 500 \Mo$ that is finally ejected and expected to grow further outside the NSC environment. There have been other merger driven proposals to assemble IMBHs in globular clusters (see for instance \cite{MillerHamilton2002,Leigh+2013,Antonini+2019} and references therein), here we focus on the gas-rich dense NSC environment and summarize below only models that include mass growth via accretion in addition to stellar dynamical processes in these particular sites.

\subsection{Properties of Observed Nuclear Star Clusters}

NSCs are observationally found to be extremely dense and massive star clusters that occupy the inner-most nuclear regions of most luminous galaxies. Detected as highly luminous compact sources, distinct from star clusters, they constitute the excess light above and beyond the inward extrapolation of the surface brightness profiles of galaxies on scales $\leq$ 50 pc. NSCs have measured central densities of $10^{6-7}\,\Mo\,{\rm pc^{-3}}$ on scales of 0.1 pc and $10^{2-4}\,\Mo\,{\rm pc^{-3}}$ on scales of 5 pc \citep{Schodel+2008,Feldmeier-Krause2017}. The nuclear regions of most, if not all galaxies seem to harbor a NSC and/or a SMBH. NSCs are abundant and appear to inhabit  $>$ 60-70\% of early and late-type galaxies \citep{Neumayer+2020}. The scaling relations of NSCs and SMBHs with their host galaxy properties are surprisingly similar - which led to the use of the more generic term of central massive object \citep{Ferrarese02}. The mass of the central SMBH and the mass of the central NSC are found to correlate with host galaxy luminosity, stellar mass, and stellar velocity dispersion. As with the empirically determined correlations between the SMBH and host bulge mass noted earlier, the origin of the correlations with the mass of NSCs is yet to be fully understood. A recent detailed review of the current status of our understanding of observed NSCs and their properties can be found in \cite{Neumayer+2020} and key formation mechanisms for NSCs are described in \cite{Antonini+2015} and references therein. 

Particularly intriguing are galaxies like the Milky Way ($M_* \sim 10^{10}\,\Mo$) in which a SMBH and a NSC co-exist. The central SMBH in the Milky Way with a mass of $\sim 4 \times 10^6\,\Mo$ is harbored in a fairly typical NSC with a surface density of $\sim 2 \times 10^5\,\Mo\,{\rm pc^{-2}}$ within the effective radius and a density of $\sim 2.6 \times 10^5\,\Mo\,{\rm pc^{-3}}$ within the inner 0.1 pc \citep{Schodel+2008,Feldmeier-Krause2017}. It has been suggested that the presence of a massive black hole might inhibit the onset of core-collapse in the NSC, causing it to expand before eventual disruption (see references in \cite{Neumayer+2020}). In addition, the tidal forces from a MBH on the NSC are likely to be significant when the radius of influence $r_{\rm inf}$ of the MBH is comparable to size of the NSC. Since $r_{\rm inf}$ scales linearly with the mass of the black hole, the impact of the MBH on stellar orbits in NSCs is more significant for more massive black holes. Independent of how this co-existing eco-system is assembled, whether via the inspiral of a NSC into a galaxy core with a pre-existing MBH or vice-versa; the integrity of the NSC and its dynamics are likely to be transformed when the MBH mass is comparable to or exceeds that of the NSC. 

Using a large sample of NSCs \cite{georgiev+16} recently explored scaling relations with their host galaxies, including cases when a MBH is harbored in a NSC. They find variations in the slope and offset of the scaling relations that depend on the morphology of the host galaxy. The following key findings from the analysis of observations by \cite{georgiev+16} are salient for the physical picture presented in this work: NSCs are found to be rarer in high-mass galaxies, which preferentially contain a central SMBH instead, suggesting that with increasing galaxy mass, processes that either prevent the formation of NSCs or those that destroy them are efficient; and for low mass galaxies, extrapolation of the relation between co-existing central BHs and NSCs with galaxy stellar mass suggests the presence of IMBHs with mass of  $10^{4-5}\,\Mo$ in them. 

\subsection{Massive black holes from runaway tidal encounters in nuclear star clusters}

\cite{stone+17} propose that as all galaxies likely harbor NSCs with masses proportional to their spheroidal mass, in those above a critical threshold stellar mass the NSCs can serve as feasible sites for the formation of an IMBH and or a SMBH from stellar remnants that could end up eventually as central BHs. They argue that runaway tidal encounters, both tidal capture and tidal disruption offer a unique growth channel for the remnant stellar mass black holes in these environments. Stellar mass black holes in NSCs, \cite{stone+17} argue, can experience runaway growth to produce IMBHs/SMBHs, driven by tidal interactions with stars in the NSC in a three-stage process each defined by stellar dynamical process that dominate. Early-on the mass growth process is extremely efficient and is driven by the unbound stars leading to supra-exponential growth. Once the BH has grown to $\sim$ 100 $M_{\odot}$, the growth switches predominantly due to the feeding of bound stars thus depleting the loss cone. In this second stage too, the mass build-up of the black hole can be extremely rapid. At later times, the growth slows down as it occurs in a diffusion limited fashion by depletion of the entire mass of the cluster core leading to eventual saturation at $M_{\rm sat} \, \sim \, 10^{5-6}\,M_{\odot} \, \propto \sigma^{3/2}\,t^{1/2}$ where $\sigma$ is the velocity dispersion of the NSC and $t$ is growth time in this final mode. Of course, for such runaway processes to operate, the NSC has to be dense, with a central density of $n_c\,\geq 10^7\,{\rm pc}^{-3}$, to enable the enhanced star-BH interaction rate required to jump-start and promote this growth channel.  Several observed low redshift NSCs appear to have the requisite densities to be viable sites for this process to proceed (see Table A1 in \citep{georgiev+16}). Remarkably with this tiered stellar dynamical processing driven model, \cite{stone+17} are able to reproduce a scaling relation between black hole mass and the velocity dispersion of the host galaxy that is close to the observed $M_{bh} - \sigma$ relation. 

\subsection{Massive black holes from core collapse and stellar collisions in nuclear star clusters} 

\cite{devecchi+09, devecchi+10} present a scenario for the formation of ($\sim 1000 \Mo$) black holes that are produced primarily via stellar-dynamical processes in the first generation of star clusters that likely assemble at $z \sim 10-20$. With conditions similar to those that are conducive to DCBH formation, a generation of early seeds, they argue, could form from a set of cascading instabilities that occur in protogalaxies with low metal content. These are expected to occur in  massive high redshift halos with $T_{\rm vir} >10^4$ K, soon after the first generation of stars - whose IMF is believed to be skewed to the high mass end - have finished their evolution.  The following additional conditions are needed to lead to seed formation: atomic hydrogen be the only efficient coolant available to cool gas; there is a background UV field from the first stars, and the gas in the halo is nearly pristine but could be polluted lightly by metals from the life-cycle of the first stars. These conditions ensure that in the next generation, dense compact star clusters form in the inner regions of these protogalaxies that in turn undergo core collapse prior to stars finishing their entire life-cycle. Following which, runaway stellar collisions can assemble a very massive star that will eventually leave behind a black hole of $1000 - 2000\,\Mo$ - an IMBH born out of stellar dynamical activity in a NSC at extremely high redshifts. Given the required conditions for the cascading set of processes, this is likely to occur only at the highest redshifts and terminate once metal cooling starts to dominate.

\subsection{Massive black holes forming from gas inflows in high redshift NSCs}

\cite{Lupi+2014} explore a scenario in which akin to the model proposed by \cite{AlexanderNatarajan2014} high redshift gas-rich NSCs are invoked as viable sites for the formation of massive BHs. Implementing a model originally proposed by \cite{Davies+2011}, in the cosmological context at high redshift, they investigate the gas driven core collapse of a dense cluster of stellar mass black hole remnants in the NSC environment. The NSC itself first forms in a pre-galactic gas disk in a dark matter halo and in order to retain the remnant stellar black holes, external gas capture is invoked to boost the depth of the potential well. Gas inflows that bring in ten times the mass of the NSC in gas - required to deepen the NSC potential to retain the stellar remnants are expected to be provided either from mergers with other gas rich haloes or from instabilities instigated in the underlying pre-galactic disk environment. Combining a cosmological simulation to track dark matter halos and baryons at high redshift, they model the formation and subsequent evolution of the NSC with a semi-analytical model and find that once relativistic core collapse is catalyzed, a final BH with masses between $300 - 1000 \Mo$, in the IMBH mass range, can form readily. In order to operate, this mechanism requires large gas inflows and a suppression of three-body interactions that might lead to the ejection of stellar remnants prior to core collapse. 

In contrast, the \cite{Davies+2011} model offers a more efficient channel that operates only in the most massive NSCs with high velocity dispersions. In NSCs with velocity dispersions between 40 - 100 km s$^{-1}$ wherein two-body relaxation times are short enough to ensure mass segregation leading to the formation of a core of stellar mass black holes. The depth of the potential well in these NSCs leads to the ejection of the majority of stellar remnants via three-body interactions or gravitational wave recoil. However, if a single or binary black hole remains un-ejected, then it is retained in the core and can grow subsequently by capturing the remaining cluster stars.

\section{The physics of continual BH formation in a dense, gas-rich NSC}

Here we present the detailed physical process that provides a robust channel to amplify and grow stellar mass remnants into a range of final BH masses in the environment of dense, gas-rich NSCs. Our proposed continual BH formation mechanism rests on the following properties of gas accretion: (i) Bondi accretion (or other accretion processes where the gas inflow rate increases with central accreting mass faster than linearly) leads to supra-exponential mass increase, diverging in a finite time; (ii) the limit imposed on the spherical accretion rate increases well above the Eddington rate with increasing optical depth approaching the Bondi rate at high optical depths and finally (iii) that spherical accretion can occur for black hole seeds embedded in gas rich NSCs. 

 \subsection{Physical Processes: supra-exponential accretion of gas by a wandering BH}
\label{ss:SEaccretion}

We begin by discussing in detail the key elements and physical processes invoked for the rapid mass growth of an initial stellar mass seed BH embedded in a gas rich NSC. Our proposed scenario is driven by supra-exponential accretion of gas with finite optical depth aided by the dynamics of the BH in the host cluster, and enabled by the evolution of angular momentum in the accretion flow due to scattering of the BH. The first point to note is that essentially any stronger-than-linear dependence of the accretion rate on the BH mass, will lead to divergent growth of the BH mass in a finite time. And this timescale for divergence can be explicitly computed as done below for the relevant case of Bondi accretion, and the shortness of this timescale compared to other relevant dynamical timescales that disrupt or render the NSC unstable is what permits the extremely rapid early growth in this environment. 

Consider the two limiting cases of spherical accretion---that of a completely transparent flow, and that of a completely opaque one. These are described by the Eddington-limited accretion solution (which assumes zero optical depth, gravity and radial radiation flux forces only, neglecting gas pressure gradients) and the Bondi solution for adiabatic index $\Gamma=4/3$ (which assumes infinite optical depth, gravity and radiation-dominated gas pressure gradients only, but no radiation flux force). 

Optically thin accretion allows radiation to escape from all regions of the flow. In the limit where gas pressure can be ignored (a consistent, but not necessary, implication of low optical depth), the balance between gravity and radiation flux pressure establishes a maximal accretion rate, the familiar Eddington limit. The Eddington luminosity is defined as:
\begin{eqnarray}
L_{\rm Edd} = \frac{4\,\pi\,G\,M_{\rm bh}\,c}{\kappa},
\end{eqnarray}
and the Eddington mass accretion rate is:
\begin{eqnarray}
\dot{M}_{\rm Edd} (\eta) = \frac{4\,\pi\,G\,M_{\rm bh}}{\eta\,\kappa\,c} \equiv\frac{\Mbh/t_{1}}{\eta_{E}}\,,
\end{eqnarray}
where $0.057\le\eta_{E}\le0.42$ is the radiative efficiency (typically, $\eta_{E}=0.1$; $\kappa$ the opacity taken to be the Thomson electron-scattering opacity and $t_{1}=\kappa c/4\pi G$ is the Eddington timescale (for $\eta_{E}=1$). Accretion at the Eddington limit results in exponential mass growth from an initial mass $M_0$ given by:
\begin{eqnarray}
 \Mbh(t)=M_{0}\exp(t/t_{E})\,
 \end{eqnarray}
 on the characteristic Salpeter timescale, 
\begin{eqnarray}
t_{E}=\frac{\eta_{E}}{1-\eta_{E}}t_{1}\,,\label{e:t_Salpeter}
\end{eqnarray}
where the typically small and often neglected, $(1-\eta_{E})$ factor in the denominator accounts for the loss of rest-mass by the escaped radiation, that is, 
\begin{eqnarray}
\dot{M}_{\rm bh}=(1-\eta_{E})\Medd. 
\end{eqnarray}
Defining $\dot{M_1} = \dot{M}_{\rm Edd}(\eta = 1) = L_{\rm Edd}/c^2$.

Briefly reviewing the Bondi picture, consider a BH of mass $\Mbh$ that accretes from a gas reservoir with density $\dinf$, sound speed $\cinf$ and opacity $\kappa$ at infinity, while moving at velocity $\vbh<\cinf$ relative to it. We assume here that the BH velocity is a small perturbation on the spherical accretion solution, and can be ignored for the purpose of estimating the general properties of the accretion flow. The opacity is assumed to be independent of the gas properties (specifically, electron scattering opacity). Gas around the BH within the accretion (or capture) radius (\citealt{hoy+39}; \citealt{edg04})
\begin{eqnarray}
r_{a}=2G\Mbh/(\cinf^{2}+\vbh^{2})\,,
\end{eqnarray}
 is dynamically bound to the BH, which is a necessary, but insufficient condition for the gas to accrete onto it. The gravitational energy of the gas that is released as radiation when it flows toward the BH can reach a distant observer only from gas emitting outside the photon trapping radius $\rtr$. The photon
trapping radius $r_{\rm trap}$ is:
\begin{eqnarray}
r_{\rm trap} = \frac{\kappa\,\dot{M}_{\rm bh}}{4\,\pi\,c} = \frac{\dot{M}_{\rm bh}}{\dot{M_1}}\,r_{\rm g},
\end{eqnarray}
where the local definition of the optical depth $\tau(r) = \kappa\,\rho(r)\,r$ is used in the trapping criterion, $v_r(r) = c/{\tau(r)}$ and the spherical continuity equation $\dot{M} = 4\,\pi\,r^2\,\rho(r)\,v_r(r)$ is assumed. 

From the outset, given the geometry under consideration, we assume that Bondi accretion with adiabatic index $\Gamma = 4/3$ appropriately describes the flow, supported also
by the analysis of \cite{Soffel1982}, without requiring us to account for the limiting effect of radiation flux pressure from a small embedded accretion disk. The smallest possible trapping radius is at the innermost stable circular orbit (ISO), which is a property of the BH's mass and of its spin relative to the gas. For a non-spinning BH, $r_{\mathrm{ISO}}=6r_{g}=6G\Mbh/c^{2}$.  However, trapping can occur at much larger distances when the accretion flow is optically thick enough so that the photon diffusion velocity upstream is slower than the inflow velocity \citep{beg78}. Therefore,
 \begin{eqnarray}
c/\tau(\rtr)=c/[\kappa\rho(\rtr)\rtr]
\end{eqnarray}
is lower than the inflow velocity \citep{beg78}, 
\begin{eqnarray}
v(\rtr)=\Mdot/[4\pi\rtr^{2}\rho(\rtr)];\,\,\,\, \rtr=\Mdot\kappa/4\pi c. 
\end{eqnarray}
The radiative efficiency at the trapping radius is:
\begin{eqnarray}
\eta_{\rm trap} = \frac{G\,M_{\rm bh}}{r_{\rm trap} c^2} = \frac{M_{\rm bh}/t_1}{\dot{M}_{\rm bh}}
\end{eqnarray}

In this case, $\rtr\gg r_{\mathrm{ISO}}$, and its location is a property of the flow. Photons emitted inside $\rtr$ are simply advected with the flow and accreted by the BH. 

Let us assume that the gravitational potential at $\rtr$ can be approximated as $\phi=-G\Mbh/\rtr$ and that the gas falls from infinity with zero energy. The radiative efficiency of the accretion flow, that is, the fraction of rest mass energy released to infinity, is then $\eta=G\Mbh/\rtr\,c^{2}$. Note that even at $r>\rtr$, the released radiation need not necessarily interact with the accretion flow upstream. If it doesn't, it will not suppress the flow, there will be no upper limit on the accretion rate, and hence on the BH growth rate as in the Bondi case. In this event, the accretion rate is set by the boundary conditions $\dinf/\cinf$ and is therefore, in principle, unbounded.  In the limit $\tau_{a}\to\infty$ the flow is adiabatic, regulated by the balance between gravity and gas pressure, which is described by the Bondi solution. 
 For now we assume that $\vbh/c_{s}\to0$  \citep{bon52} with radiation-dominated gas pressure (adiabatic index $\Gamma=4/3$), and the relevant Bondi accretion rate is:
\begin{eqnarray}
\Mdot_{B}(\Gamma=4/3)=\frac{\pi}{\sqrt{2}}r_{a}^{2}\rho c_{\infty}=\pi r_{a}^{3/2}\sqrt{G\Mbh}\dinf\:.\label{e:MdotB}
\end{eqnarray}
The Bondi accretion rate equation can be cast in a form similar to that of the Eddington limit (Eq.~2) as done below:
\begin{eqnarray}
\Mdot_{B}=\frac{\sqrt{2}\pi\tau_{a}G\Mbh}{\kappa\cinf}\equiv\frac{\Mbh/t_{1}}{\eta_{B}}\,,\label{e:MdotBtau}
\end{eqnarray}
where 
\begin{eqnarray}
\eta_{B}=\frac{2^{3/2}}{\tau_{a}}\frac{\cinf}{c}\equiv\frac{\eta_{B,1}}{\tau_{a}}\,,
\end{eqnarray}
is defined in terms of $\tau_{a}=\kappa\dinf r_{a}$, the optical depth on the scale of the accretion radius, evaluated with the asymptotic density $\rho(r_{a})\sim\dinf$. It then follows that (Eq.~10)
\begin{eqnarray}
\rtr=\left(\frac{\tau_{a}}{2^{5/2}}\frac{\cinf}{c}\right)r_{a}\,,
\end{eqnarray}
and that the condition for adiabatic flow ($\rtr>r_{a}$) is:
\begin{eqnarray}
 \tau_{a}>2^{5/2}(c/\cinf).
 \end{eqnarray}
Note that a similar condition was derived by \citet{beg78}, $\tau_{a}>(2^{3/2}/\sqrt{3})(c/\cinf)$. For an optically thick flow where the radiation is trapped and accreted by the BH, $\dot{M}_{\rm bh} \simeq \Mdot_{B}$ (cf Eq. \ref{e:t_Salpeter}). 

Unlike the Eddington accretion rate which scales $\propto\Mbh$, and therefore results in exponential growth, wherein the BH mass diverges only as $t\to\infty$, the Bondi accretion rate scales as $\propto\Mbh^{2}$, and its solution has a supra-exponential  accretion rate, where the mass diverges in a finite time $\tinf=\eta_{B}(0)t_{1}$,
\footnote{We note here that supra-exponential growth is not limited to the $\tau_{a}\gg1$ case of Bondi accretion. Any stronger than linear ($\Mdot_{\bullet}\propto\Mbh$) dependence suffices. For example, consider the scaling,
\begin{eqnarray}
\Mdot_{\bullet}=\Mbh^{1+\alpha}/(M_{0}^{\alpha}t_{\alpha}),
\end{eqnarray}
where $\alpha>0$, $t_{\alpha}$ is the accretion timescale and $M_{0}$ is the initial mass. The growth solution is then 
\begin{eqnarray}
\Mbh(t)=M_{0}/\left[1-t/\tinf\right]^{1/\alpha},
\end{eqnarray}
where $\tinf=t_{\alpha}/\alpha$ is the time to divergence, which as expected goes to infinity in the limit $\alpha\to0$ (Eddington).  Physically, the divergence implies that the BH will exhaust its mass reservoir by time $\tinf$, and from that point on, the accretion will be entirely supply-limited.}
where $\eta_{B}(0)$ is evaluated with $\tau_{a}(M_{0})$,
 \begin{eqnarray}
\Mbh(t)=M_{0}\left/\left[1-t/\tinf\right]\right.
\end{eqnarray}

The $\mathrm{d}\Mbh/\mathrm{d}t\propto\Mbh^{2}$ scaling and rapid mass divergence of the high optical depth limit is that of the simple radiation-less, wind accretion rate, 
$\dot{M}_{\rm bh}=\pi r_{a}^{2}(\Mbh)\bar{\rho}v$. This is the appropriate rate when radiation pressure back-reaction is irrelevant. 

Radiation produced interior to the trapping radius is advected into the BH, as the high local optical depth in this region renders the outward diffusion of photons to be much slower than accretion inward, therefore the luminosity that escapes outward to infinity $L_{\infty}$, has an effectively lowered radiative efficiency $\eta_{\rm trap}$ and does not exceed $L_{\rm Edd}$ while permitting supra-exponential accretion \citep{beg78}. As outlined in \citep{AlexanderNatarajan2014}, the corresponding effective radiative efficiency for the radiation that escapes to infinity is:
\begin{eqnarray}
\eta_{\rm trap} = \frac{G\,M_{\rm bh}}{{\rm max}\,(r_{\rm trap},r_{\rm ISO})\,c^2} = {\rm min}\,(\frac{r_g}{r_{\rm trap}},\frac{r_g}{r_{\rm ISO}}),
\end{eqnarray}
where $r_{\rm ISO}$ is the inner-most stable circular orbit. And the resulting luminosity from spherical accretion $L_{\rm S}$ which can never exceed the Eddington rate, irrespective of the mass accretion rate, is given by:
\begin{eqnarray}
L_{\rm S}  = \eta_{\rm trap}\,\dot{M_{\rm S}}\,c^2 ={\rm min}\,({\frac{\dot{M}_1}{\dot{M}_{\rm S}}},{\frac{r_g}{r_{\rm ISO}}})\,\dot{M_{\rm S}}\,c^2\,\\
\leq\,{\rm min}\,({L_{\rm Edd}},{\frac{r_g}{r_{\rm ISO}}}\,\dot{M_{\rm S}}\,c^2)\,\leq\,L_{\rm Edd}.
\end{eqnarray}
\cite{beg78} estimated $L_{\rm Bondi} \simeq 0.6\,L_{\rm Edd}$, taking into account the support against gravity provided by the escaping radiation. And the mass accretion rate is then:
\begin{eqnarray}
\dot{M}_{\rm Bondi} = \frac{\pi\,r_{\rm a}^2\,\rho_{\infty}\,c_{\infty}}{6\sqrt{2}},
\end{eqnarray} 
where the mass accretion rate above is suppressed by an extra-factor of $1/6 \sim (1 -0.6)^2$ with respect to the standard Bondi rate (see Eq.~11) due to the lowered value of  $L_{\rm Bondi} \simeq 0.6\,L_{\rm Edd}$. The supra-exponential mass growth rate for the initial BH seed of mass $M_{\rm i}$ is:
\begin{eqnarray}
M_{\rm bh}(t) = \frac{M_{\rm i}}{(1 - \frac{t}{t_{\infty}})}
\end{eqnarray}
where the growth divergence time-scale $t_{\infty}$ is given by:
\begin{eqnarray}
t_{\infty} = \frac{3}{\pi\,\sqrt{2}}\,\frac{c_{\infty}}{G^2\,M_{\rm i}\,\rho_{\infty}}
\end{eqnarray}

\subsection{Timescales relevant to BH dynamics in the host NSC}

The rapid mass amplification provided in the environment of a typical gas-rich NSC comprises three stages, the first is when the initial seed BH with mass comparable to or slightly larger than that of a typical cluster star can be treated as test particle in equipartition with the stars that it scatters off.  The BH wanders around as it attains dynamical fluctuation-dissipation equilibrium with the stars in the NSC via two-body interactions. At this point the BH is in equipartition with the stars and wanders randomly (Stage 1). During these excursions it gains mass via Bondi accretion (defined and shown as Stage 2 in the schematic Fig.~1) followed by a  transition to the next growth stage when the mass of the BH has grown to be large enough that its motion is damped and it moves toward settling down at the center of the NSC (defined as Stage 3). Once the damped motion of the BH ceases,  it ends up as a stationary BH at the center of the NSC. Subsequent growth is likely to occur in an Eddington limited fashion provided there is gas is available in the reservoir. The final mass is expected to be the IMBH mass range.

For our picture of early mass growth in Stages 2 and 3 to work, the wandering BH fed by wind accretion requires a stable NSC core environment. Therefore, the dynamics of the embedded wandering BH needs to be followed in the larger context of the host NSC and its relevant evolutionary timescales. The BH mass build-up timescale can then be computed given the properties of the accretion flow. 

Therefore, first we need to compute and compare the various characteristic timescales for dynamical processes that operate in the NSC. The relevant processes and their characteristic timescales to consider are:  the two-body relaxation timescale evaluated usually at the half mass radius $t_{\rm rh}$; the collisional time-scale $t_{\rm coll}$ for stellar collisions; and the evaporation timescale $t_{\rm evap}$ for the cluster. Here, we argue that the vector resonant relaxation timescale $t_{\nu {\rm RR}}$ is an important new determinant of the fate of the BH and its growth cycle as this is the timescale on which the angular momentum of the accreting BH is randomized. To calculate these timescales explicitly, the Plummer model is used to model a fiducial NSC. It is the ratio of the mass divergence timescale (Eqn. 24) to the two-body relaxation timescale that determines the mass growth factor of the BH in Stage 2. The growth process can stall prior to the end of the BH's damped random walk (Stage 3) through the nucleus of the NSC. Termination of the early growth process can occur and as we show later, this will lead to the production of lower final BH masses.

The initial seed BH in this scenario is expected to be provided by core-collapse, from Pop~II progenitors, and is expected to be ${\cal O}(5-10)$ times more massive than the mass of a typical cluster star, taken to be $\sim\,1\,\Mo$ in its birth cluster.  In previous work, examining the high redshift case, \cite{AlexanderNatarajan2014} adopted initial conditions relevant for PopIII remnants and the expected gas content of  NSCs that form at $z \sim 10 - 15$ with guidance from cosmological simulations \cite{Wise+19}. For models considered here, we use the observationally determined parameters provided in Table A1 for the fiducial NSC properties \citep{georgiev+16}.

The constituents of gas rich NSCs at later epochs considered here are modeled as follows: the mass distribution of the inner region is modeled as a spherical constant density core of a Plummer model with a total mass in stars and gas of $M_{\rm NSC}$ with radius $R_{\rm NSC}$, and with 1D velocity dispersion $\sigma$ assuming that the velocity distribution is Maxwellian. For an NSC that contains $\Ns$ stars of mass $\Ms$ each and a single BH initially of mass $\Mbh$, we define the following quantities, the mass ratio $Q=\Mbh/\Ms$; cluster velocity $V_{\rm NSC}^{2}=GM_{\rm NSC}/R_{\rm NSC}$ and the cluster angular velocity $\Omega_{\rm NSC}^{2}=R_{\rm NSC}^{3}/GM_{\rm NSC}$. Further, we assume that initially the mean energy of the wandering BH is given by, $E_{\rm bh}=3\Ms\ss^{2}$ is in equipartition with that of the stars, parametrized by the velocity dispersion for the stars $\ss$. The equations governing the motion of the BH define the excursions from the center driven by scattering off the stars in the potential of the constant density core. These represent small perturbations from the center and are aptly described at the beginning by a 3D harmonic oscillator during Stage 1 of the growth cycle, 
\begin{eqnarray}
 \ddot{\boldsymbol{r}}=-\Omega_{0}^{2}r\hat{\boldsymbol{r}},
 \end{eqnarray}
where $v^{2}/2+\Omega_{0}^{2}r^{2}/2=E_{\rm bh}/\Mbh$, and the fundamental frequency is given by, $\Omega_{0}^{2}\sim{\cal O}(\Omega_{NSC}^{2})$. Equipartition then implies that the BH's rms 3D velocity is $\vbh^{2}=3\sbh^{2}=(3/Q)\ss^{2}$ and its rms 3D displacement from the center is $\dbh^{2}=(3/Q)\ss^{2}/\Omega_{\rm NSC}^{2}$. Initially, assuming that the gas and stars both share the cluster's velocity dispersion, $\sigma^{2}\sim V_{\rm NSC}^{2}$, it then follows that the velocity of the BH is given by $\vbh^{2}\sim V_{\rm NSC}^{2}/Q$  and the wandering distance is given by $\dbh^{2}\sim R_{\rm NSC}^{2}/Q$ that is, the BH's typical velocity is lower than the typical velocity of stars in the NSC potential, and its excursions away from the center are confined to the core, where the density is nearly constant.

Two-body interactions of the roving BH with the cluster stars will lead to the establishment of a statistical equilibrium between the energy and angular momentum of the BH and the stars on the two-body relaxation timescale and will change the orbital energy and angular momentum of the BH by order unity around their equipartition values on the timescale given by:
\begin{eqnarray}
t_{\rm rh} \simeq \frac{0.9\,N_*}{[s^2\,\Omega_{\rm NSC}\,log (0.28\,N_*)]}
\end{eqnarray}
which is very short for the typical dense NSC parameters \citep{giersz+94}. The equilibrium driven by the balance of the two-body scattering (fluctuations) and dynamical friction (dissipation) is that of equipartition if the velocity distribution of stars in the NSC is assumed to be a Maxwellian as done here \citep{chatterjee+02}. In this event, exact expressions for the velocity and wandering distance for the BH in the Plummer potential \citep{chatterjee+02} can be calculated, and are given by:
\begin{equation}
\vbh^{2}=\frac{2^{5/2}}{3}V_{c}^{2}/Q\,,\,\,\,\dbh^{2}=\frac{2}{3}R_{c}^{2}/Q.\label{e:vbhdbh}
\end{equation}

The internal dynamics of the NSC determines its stability and lifetime. The stars will eventually destroy each other by physical collisions and the collisional destruction time for stars with typical mass of $\sim 1 \Mo$ and size of a solar radius is given by:
\begin{equation}
t_{\mathrm{coll}} \simeq 2.1 \times 10^{12} (\frac{n}{10^6\,{\rm pc^{-3}}})^{-1}\,(\frac{V_{\rm NSC}}{30\,{\rm km s^{-1}}})
\end{equation}
It will however take multiple encounters for destruction by collisions, and this is expected to occur on a timescale of the order of $10^{10-11}$ yr for the typical values for NSCs adopted here \citep{murphy+91}. Even if the NSC manages to remain undisrupted by collisions, it will eventually evaporate on a timescale of: 
\begin{eqnarray}
t_{\mathrm{evap}}\sim100t_{rh} \sim 10^{10}\, {\rm yr} 
\end{eqnarray}
estimated once again for the typical values adopted here \citep[e.g.][]{binney+08}.

Dissolution of the cluster via two-body ejections of stars occurs on a significantly longer timescale and can be ignored here \citep[e.g.][Sec 7.5.2]{binney+08}. and the ejection of the BH itself by two-body interactions is suppressed by the mass ratio $1/(1+Q$) and is therefore unlikely \citep[e.g.][Eq. 16.3]{heggie+03}.

As the wandering BH accretes mass via subsonic wind accretion in Stage 2, the motion will subsequently be damped as it transits to Stage 3. As rapid mass growth progresses, once the system is no longer in equipartition, which occurs when the initial seed as grown sufficiently to the value $M_{\rm eq}$ the motion will start to decay. The equation of motion that best describes the motion at this juncture is that of a damped 3D harmonic oscillator, with damping terms on the accreting BH provided by (i) the drag force of the non-rotating pressure supported gas - parameterised using an accretion coefficient $\gamma_{\rm a}$ below and (ii) the drag provided by dynamical friction - parameterized using the coefficient $\gamma_{\rm df}$ below. The motion thereafter is described by the following equation of motion:
\begin{eqnarray}
\ddot{\boldsymbol{r}}= -2(\gamma_{\rm a}+\gamma_{\rm df})\dot{\boldsymbol{r}}-\Omega_{0}^{2}r\hat{\boldsymbol{r}},
\end{eqnarray}
according to which the BH will come to a halt as it attains a terminal mass of $\sim 100 - 1000 \Mo$ and settles down at the center of the NSC. At this juncture, post Stage 3, growth will continue via accretion that is limited only by the availability of gas. And in principle, this now central BH can easily grow to a final mass between  $10^{4-5}\,\Mo$. The final mass depends on the gas mass available, and potential replenishment of the gas reservoir which depends on the location of the NSC with respect to the galaxy center as well as the probability of a galaxy merger. 

As the BH becomes massive as it accretes non-rotating gas (relative to the cluster center), it slows down and sinks to the center, it can no longer be treated as a test particle, and it begins affecting the dynamics and evolution of the cluster. If its accretion rate is faster than the two-body relaxation rate, as must occur eventually when the accretion diverges, equipartition can no longer be maintained, and the BH becomes ever slower due to damping. As the failure of equipartition starts to occur the extrapolation of the initial conditions and dynamics still remain reliable and this permits estimation of the terminal BH mass at Stage 3.  We estimate a range of values for specific NSC parameter choices. The terminal mass at the end of Stage 3 is dictated by the race between the decay of the BH orbit, and the increase in the captured angular momentum with BH mass. At this juncture the accretion radius $r_{\rm a}$ is no longer much smaller than the BH orbit and we use this as the limiting criterion to compute the terminal mass at the end of Stage 3.

 \subsection{Physical Processes: Angular momentum considerations}

In our proposed picture, as noted above the accelerated motion of the light BH relative to the stars is unavoidable, in fact, there is no physical mechanism that could keep the BH stationary in the early stages. This motion is the challenge, as on the one hand it permits jump-starting supra-exponential early growth while it also induces angular momentum in the accretion flow on the BH. In this picture, we assume that the net angular momentum of the NSC gas is zero and this is justified by making the case that during the formation process at high redshift gas is feed into the NSC core from filaments that effectively ensure randomization of the angular momentum of the gas. This assumption is supported by the cosmological simulations of the formation of the first baryonic structures (see for instance \cite{Regan+2009,Choi+2013,Woods+2017,Wise+2019}). However, what mitigates this problem is that the motion of the BH is entirely {\bf random}. and that the motion decays following rapid accretion after which the BH becomes stationary. We present below a set of arguments previously provided in \cite{AlexanderNatarajan2014}, to show that the angular momentum during this early growth stage is in fact very low or nearly zero during the wandering motion of the BH. While equipartition holds, a BH with mass $M_{\rm bh}\,\sim Q\,M_*$ on a typical orbit that is comparable to the wandering radius $\Delta_{\rm bh}$ will tend to accrete mass with low angular momentum due to the cancelation of the induced velocity and density gradients. This is expected in Stage 2 of the process outlined in this work. The random nature of the motion prevents the accumulation of angular momentum in the accretion flow particularly during the crucial early stages. In terms of physical processes that help curtail the angular momentum of the accretion flow, the process of resonant relaxation \cite{rauchtremaine96, rauchingalls98} plays an important role and provides a rapid relaxation mechanism for the angular momentum. However, the ultimate solution for the angular momentum problem is for the BH to stop accelerating and remain stationary in a spherical flow as required for the Bondi solution, which we find is the ultimate fate once the excursions from the center are finally damped in Stage 3.

Wind accretion while the BH is roving can proceed at rates exceeding the Eddington limit if the specific angular momentum in the wind flow, $j_{w}$, is lower than that of a plunge orbit, $j_{0}=4G\Mbh/c$. In that case the gas can flow directly into the BH without encountering an angular momentum barrier and forming an accretion disk \citep{illarionov+01}. As outlined in \cite{AlexanderNatarajan2014}, here too we assume that the gas in the NSC is pressure-supported and therefore has little angular momentum. However, we expect velocity and density gradients to be induced by the random motion across the capture cross-section $\pi r_{a}^{2}$ that will generate angular momentum $j_{w}=(\mathrm{d}\vrel/\mathrm{d}r)|_{\rbh}r_{a}^{2}(\rbh)/4>j_{0}$, where $r_{\rm bh}$ is the distance of the BH from the cluster center. The typical
velocity gradient for rotationally supported gas motion, a conservative upper limit for the pressure supported gas considered here, is $|\mathrm{d}\vrel/\mathrm{d}r|\sim|\mathrm{d}v_{\phi}/\mathrm{d}r|<\Omega_{0}$. The BH's own accelerated motion around the cluster center also creates a velocity gradient across $r_{a}$, whose magnitude is similarly $|\mathrm{d}v_{\mathrm{\phi}}/\mathrm{dr|}\sim\Omega_{0}$. We note that the magnitude of the angular momentum in the accretion flow due to orbital acceleration with frequency $\Omega_{0}$ can be estimated by considering the accretion cross-section as a thin circular disk rotating around its diameter at that frequency, so that $j_{w}=I\omega$, where $I=MR^{2}/4$ is the moment of inertia of a disk about a diameter \cite{frank+02}. This translates to $j_{w}^{\mathrm{acc}}\sim\Omega_{0}r_{a}^{2}/4$. It then follows that $j_{w}^{\mathrm{acc}}/j_{0}\sim(\pi^{4}/9\cdot2^{5/4})(\Mbh/M_{\rm NSC})(c/V_{\rm NSC})$, where the virial sound speed $c_{s}^{2}=(3/\pi^{2})GM_{\rm NSC}/R_{\rm NSC}$ was adopted. For the fiducial cluster model considered here, typically $j_{w}^{\mathrm{acc}}/j_{0}>1$. 

As the BH is continuously driven to sub-equipartition motion (lower accelerations) by the accretion of non-rotating gas, two-body scattering by the stars in the cluster can drive it back to equipartition (and higher accelerations) only as long as the growth rate is slower than the relaxation rate. Assuming that supra-exponential growth can occur even with $j_{a}/j_{\mathrm{ISO}}\sim$ few, then the BH begins to decouple dynamically from the cluster early in its growth during Stage 2. From that point on it is expected that the BH will decelerate, and angular momentum will grow more slowly. Together with the suppression of angular momentum provided by resonant relaxation, it is estimated that the supra-exponential growth to $\sim 10^{2-3}\, M_{\odot}$ is possible as the BH reaches and settles down in the center of the NSC. The assumption that $j_{a}/j_{\mathrm{ISO}}\sim$ few, is compatible with supra-exponential accretion is based on the fact that the initial azimuthal velocity of the flow at the accretion radius, $v_{a}=j_{a}/r_{a}$ is typically tiny for NSC models, $v_{a}/c_{s}\sim{\cal O}(M_{\rm bh}/M_{\rm NSC})$.  In that case, nearly 100\% of the thermal distribution of gas that is flowing into the BH has $j<j_{\mathrm{ISO}}$ and can plunge directly into the BH, while only a fraction $\sim v_{a}/c_{s}\lesssim 10^{-3}$ must circularize. 

In this case, the captured gas will initially accumulate around the BH with some angular momentum $\boldsymbol{j}_{w}$, as determined by the nearly static boundary conditions at $r_{\rm a}$. However, assuming that the mean relative rotation between the gas and stars is small (as is plausible if the stars were formed from the cluster gas), then after a two-body relaxation timescale $t_{\rm rh}$, the BH will reverse its motion relative the gas, and the direction of $\boldsymbol{j}_{w}$ will correspondingly also reverse \footnote{The difference between the a velocity gradient due to the BH's acceleration relative to non-rotating gas (i.e. the same mean angular velocities for the BH and the gas, $\left\langle \omega_{\rm bh}\right\rangle -\left\langle \omega_{g}\right\rangle =0$) and that due to true rotation ($\left\langle \omega_{\rm bh}\right\rangle -\left\langle \omega_{g}\right\rangle \neq0$) is that only in the former case scattering can cancel out the angular momentum of the accretion flow.}.

At that point, fresh counter-rotating gas will mix with the stalled accretion disk, and it will quickly drain into the BH. Assuming that all the captured mass has $j_{w}>j_{0}$, and that the accretion disk is inviscid and completely stalled, the ratio between the maximal disk mass that can accumulate in a relaxation time and the mass of the BH is (assuming the high accretion rate limit $\tau_{a}\gg1$ \citep{AlexanderNatarajan2014},
\begin{equation}
\frac{\max M_{d}}{\Mbh}=\frac{\tau_{a}\Medd t_{\rm r0}}{\Mbh}=\tau_{a}\frac{t_{\rm r0}}{t_{E}}\simeq\frac{t_{\rm r0}(0)}{t_{\infty}}\,,\label{e:maxMd}
\end{equation}
where $t_{\rm r0}$ is the central two-body relaxation time (Eq. \ref{e:tr}) and $\tinf$ is the mass divergence timescale (Eq. \ref{e:tinflim}).
For the cluster models considered here (Table 1), $c_{s}^{2}\simeq\sigma^{2}$ and $\max M_{d}/\Mbh\sim1.5(\mu-1)\sigma Q/[\eta c\log(0.1N_{\mathrm{*}})]\ll1$. 
This implies that (1) The gas will reach the BH without fragmenting, since the gravitational stability criterion is satisfied, $\max M_{d}/\Mbh<[H/R]_{r_{a}}=c_{s}/v_{k}(r_{a})=\sqrt{2}$, where $H/R$ is the disk scale length and $v_{k}$ is the Keplerian velocity; (2) Even if the accretion flow is stalled on timescales $<t_{\rm r0}$, the disk will be drained many times over the mass divergence timescale, and therefore the time-averaged accretion rate will remain high; and (3) The frequent reversal of $J_{w}$ will keep the BH spin, and therefore the radiative efficiency, low, which will in turn facilitate a high mass accretion rate in the critical initial growth phase (see the schematic of this scenario in Fig.~S1 in \cite{AlexanderNatarajan2014}). 

We also take guidance from findings of previous work wherein self-correction in Bondi-Hoyle accretion that occurs in the presence of a small inhomogeneity either a density gradient or a more generalized velocity gradient (as studied by \cite{daviespringle80,ruffert99}), leads to the near cancelation of the net angular momentum accreted. This also ensures that the angular momentum in the flow actually drops sharply during the critical early stages of growth between $M_{\rm i}$ and $M_{\rm eq}$ permitting unimpeded growth. The generation of density and velocity gradients in the accretion flow and details of the capture process are presented in detail in \cite{AlexanderNatarajan2014}, here we distill the key points to make the case that during the early stage of subsonic wind accretion onto the roving BH, the angular momentum barrier does not stifle growth. 

Our picture of a wandering BH accreting in an inhomogenuous medium with both density and velocity gradients is yet to be numerically studied in simulations, however, ballistic wind accretion in the hypersonic limit wherein gas pressure can be neglected has been studied and we take guidance from these treatments. Work by \cite{daviespringle80}, 2D calculations show that the cancelation of angular momentum we rely on here, does occur in the presence of both density and velocity gradients in the medium. In fact, they recover the same behavior to first order in the presence of these inhomogenities as the homogenous case. The 3D case was investigated with hydrodynamical simulations by \cite{liviosoker+86}, where they studied mass and angular momentum capture in the presence of density gradients and find that the mass capture rate is roughly the same as for the case of homogenous wind accretion though the capture efficiency is low. \cite{ruffert99} studied the mass capture in flows with both density and velocity gradients and found that it was high and that the efficiency of angular momentum capture though depended strongly on the details of the flow. Here, we assume an average value from these studies for the typical angular momentum capture efficiency of $\sim 1/3$.
The specific angular momentum in the accretion flow is given by:
\begin{eqnarray}
j_{\rm a} = \frac{1}{4}\,({\frac{\rm d log \rho}{\rm d log r}}-{ -2 ({\frac{3u_{\rm a}^2-1}{u_{\rm a}^2+1}})\,{\frac{\rm d log v}{\rm d log r}}})\,\Omega_{r_{\rm bh}}\,{r_{\rm a}^2}
\end{eqnarray}
where $u_a$ is the velocity of the BH at the accretion radius, and both velocity and density gradients as well as the angular momentum are computed along the orbit of the BH on its oscillations away from the  NSC center. The density gradient can be computed explicitly for the Plummer model, for which it is dimensionless and always negative:
\begin{eqnarray}
{\frac{\rm d log \rho}{\rm d log r}} = -5 \frac{(r/R_{\rm NSC})^2}{1+(r/R_{\rm NSC})^2}
\end{eqnarray}
The fiducial parameters for which these density and velocity gradient terms cancel and give $j_{\rm a} = 0$ suggest that the angular moment barrier can be circumvented in the early stages.

\section{ Results: Estimating BH growth in a NSC}
 
 Adopting a Plummer model for a fiducial NSC, parametrized by range of characteristic masses and radii typical of moderate density NSCs shown in Table~1, we compute the BH mass growth factors for a divergence timescale $t_{\infty} = 2 \times 10^7$ yr, for two values of the initial BH seed mass $M_{\rm i}\,=\,5 \Mo$ and $M_{\rm i}\,=\,10 \Mo$  in Stage 1. We find that the dynamics in these moderately dense clusters are reasonably slow with typical cluster velocities ranging from $13.1 - 18.8 {\rm km}\,{\rm s}^{-1}$; the computed half-mass relaxation timescales ranging from $1.7 - 7.4 \times 10^8$ yr are longer than $t_{\infty}$; and the calculated evaporation and collisional destruction timescales are extremely long ($> 10^{10}$ yr). The vector resonant relaxation timescales ranging from $ 1 - 1.7 \times 10^6$ yr are short enough to be relevant as discussed for slowing down the growth of an accretion disk by the randomization of the BH orbit, in fact, $t_{\nu RR} \sim (1/20) t_{\infty} - (1/10) t_{\infty}$, yielding angular momentum suppression by resonant relaxation by factors $f$ ranging from 5 - 10 where $f = \sqrt{{t_{\infty}}/t_{\nu RR}} $.  For the models explored here, we find that the final BH mass at the end of Stage 2 ranges from $M_{\rm bh} \sim 25 - 100 \Mo$.
  
 \begin{figure} 
 \includegraphics[width=1.0\columnwidth]{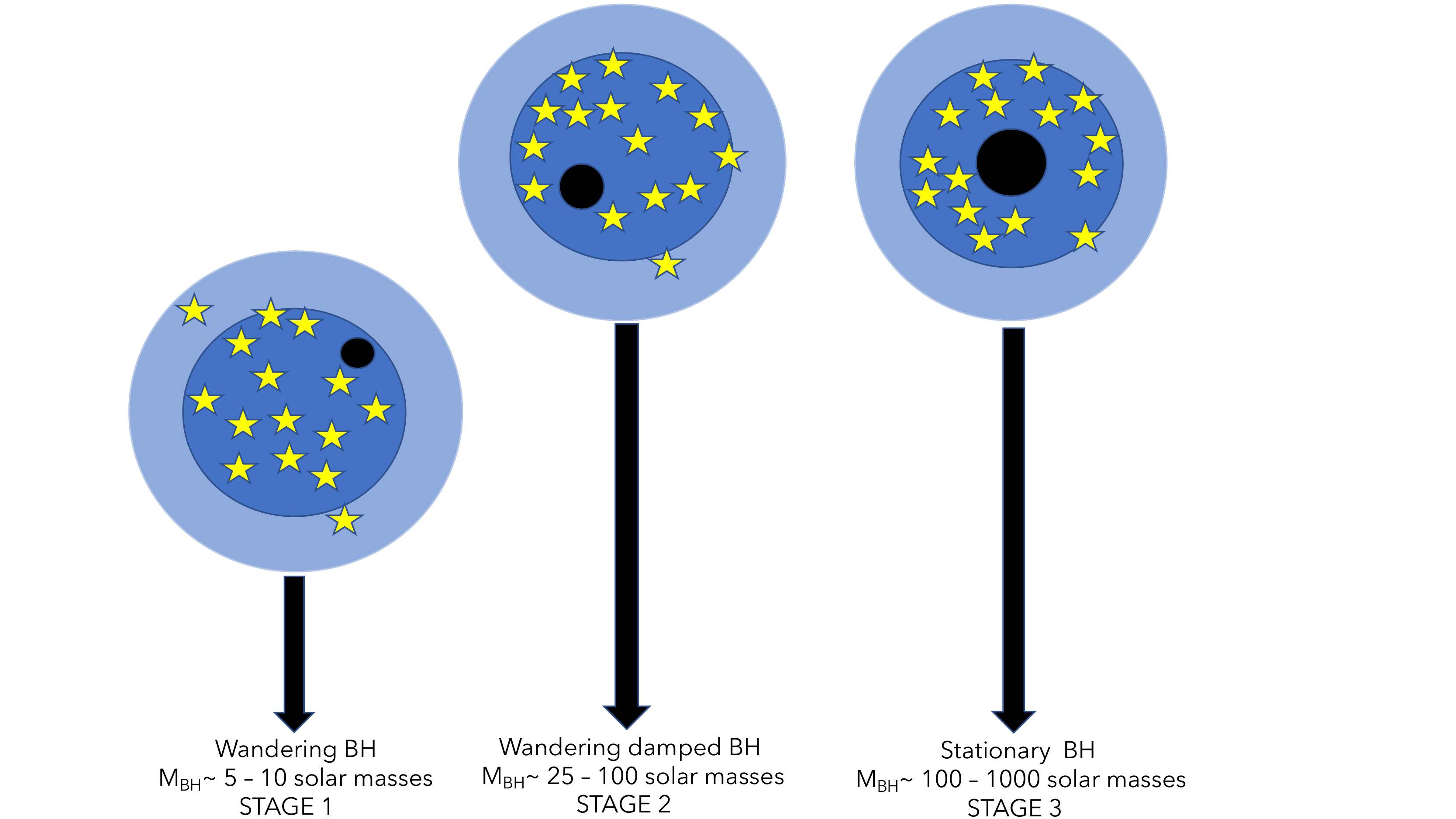} 
 \caption{Schematic figure showing the range of BH masses that could be grown in the centers of dense, gas-rich NSCs. Starting with a seed stellar remnant with mass ranging from $5 - 10 \Mo$, that is in equipartition with stars in the NSC, it experiences rapid growth via wind-fed accretion while wandering around the core of the NSC. As it gains mass, it can no longer to be kept in equipartition with the stars and gas and starts to get dynamically decoupled from the stars. This occurs when it has grown to $25 - 100 \Mo$ (Stage 2). At this juncture, it starts to decelerate and make its way to the center and starts accreting as a stationary, central source and it could grow in this phase to a final mass of $100 - 1000 \Mo$ (Stage 3). Depending on the availability of gas, it could keep growing up to
 a terminal mass of $10^5 \Mo$ at which point all the available gas would be close to depleted.}
 \end{figure}
  
 So out to this point in the very early growth the wandering BH on orbits of the scale of the wandering radius $\Delta_{\rm bh}$ will accrete with very little angular momentum. Past this Stage 2, we argue that the randomizing of the angular momentum effected by vector resonant relaxation ensures only very low residual angular momentum in the accreted gas. Vector resonant relaxation will be effective as long as the BH is conducting random excursions away from the NSC center till the assumption of equipartition ceases to be valid. As argued here and as derived in detail in \cite{AlexanderNatarajan2014}, the most severe constraints are in the very early growth stages and once, the wandering BH has grown to a final mass that is $\sim (20 - 100) \times M_{\rm i}$ its initial seed mass, the exponential growth phase terminates. This leads to masses at Stage 3 ranging from $100 - 1000 \Mo$ IMBHs and further growth will occur at rates entirely dictated by the available gas supply in the inner regions of the NSC. In gas rich NSCs considered here, these BHs can easily grow to into IMBHs with masses of $10^{4-5}\,\Mo$, if accretion proceeds in an Eddington limited fashion or even at sub-Eddington rates thereafter. Therefore, for modest redshift NSCs ($3 < z< 1$) there is ample time to grow these to the IMBH masses currently detected in low-mass galaxies.  So, in effect we show that the combination of accretion physics coupled with the dynamics of NSCs offers a very conducive environment to produce BHs ranging in mass from $100\,\Mo$ to $\sim 10^{4-5}\,\Mo$. Therefore, gas-rich, dense NSCs can effectively serve as incubators for the formation of IMBHs throughout cosmic time.

We note that the above description of the accretion and dynamical evolution of the black hole in the NSC environment rests on two key simplifying assumptions, (i) that the gas in the NSC has net zero angular momentum or at the very least very low angular momentum and (ii)  that radiative feedback from the growing black hole does not impact the accretion flow. We briefly justify these simplifying assumptions. The progression of mass growth will be inevitably impeded if the gas has angular momentum and radiative feedback is significant. Here we assume that the gas in the NSC is pressure supported and therefore has little angular momentum. This assumption is justified from cosmological simulations of high redshift \cite{Regan09} and \cite{Wise+2019} protogalactic environments, since the gas falling into collapsed halos flows in from all directions and causing randomization and cancellation of the angular momentum on average. In our picture, we assume that the NSCs form in such halos.  At present no simulations that are tailored to the mimic physical scenario that we have sketched out in this paper exist. Tackling the issue of radiative feedback and its impact on the roving black hole during the wandering stage, there are no apt simulations or numerical treatments currently available that are salient to the picture presented here. Therefore, we take guidance from the numerical studies by \cite{ParkRicotti2011,ParkBogdanovic2017,Pacucci+2017} of the impact of radiative feedback on the mass accretion rate studied in slightly different contexts. \cite{ParkRicotti2011} explore spherical accretion onto IMBHs in the first galaxies to form at high redshift   from a uniformly dense gas with zero angular momentum and extremely low metallicity. Performing idealized 1D and 2D simulations, they explore how X-ray and UV radiation emitted near the black hole regulates the gas supply and hence the accretion rate. In a suite of simulations investigating a range radiative efficiencies, black hole masses, densities and sound speeds of the ambient gas they calculate how radiation modifies the Bondi solution. They find that the average accretion rate is suppressed compared to the Bondi rate but oscillates due to X-ray and UV photo heating, however, the effects are minimal for gas densities $n_{\rm gas} \leq ({10^{5} - 10^{6}})\,{\rm cm^{-3}}$. Incidentally, this is the gas density range that we assume for our fiducial NSC while computing the mass accretion history. While their studies focused on zero angular momentum gas, they speculate that inclusion of a small, non-zero angular momentum for the gas would likely alter the time-dependent behavior of the accretion rate.  In a slightly different set-up, following black hole mergers precipitated by their host galaxy mergers, \cite{ParkBogdanovic2017}, study the efficiency of dynamical friction in the presence of radiative feedback from a black hole moving through a uniform density gas using idealized 2D simulations. For black holes masses $< 10^{7} \Mo$ they find a weakening of the dynamical friction force therefore delaying the in-spiral of the secondary black hole to the center. The implications of this work for our set-up suggest that the damped oscillation stage (Stage 2) is likely to last longer if radiation feedback is taken into account. In other idealized simulations of a central growing seed black hole, performed by \cite{Pacucci+2017}, once again not the exact physical picture explored here, they find that there exists an optimal parameter space for growth by stable super-Eddington accretion that is defined by the black hole mass and the local gas density. Extremely efficient steady growth at super-Eddington rates can be expected in the region where the ambient gas density ranges from $10^{6} - 10^{8}\,{\rm cm^{-3}}$ for black hole masses between $10$ - few hundred $\Mo$ taking radiative feedback into account (see top panel of Figure 1 in cite{Pacucci+2017}). Though Pacucci et al. (2017) considered central, stationary black holes, the region that permits stable super-exponential accretion is precisely the region where the wandering BH presented in our picture is situated.

\begin{table*}
\centering
\begin{tabular}{lcccccccccc}
\hline
        & $M_{\rm NSC}$ & $R_{\rm NSC}$ & $Q$ & $V_{\rm NSC}$ & $v_{\rm bh}$ & $\Delta_{\rm bh}$ & $t_{\rm rh}$ & $t_{\nu {\rm RR}}$ & $t_{\rm evap}$  \\ \hline
\\
Model A1Aa & $2 \times 10^4 \Mo$ & $0.25\,{\rm pc}$ & 5 & $18.8\,{\rm kms^{-1}}$ & $11.5\,{\rm kms^{-1}}$ & $0.08{\rm pc}$ & ${1.7 \times 10^8\,{\rm yr}}$ & ${1.8 \times 10^6\,{\rm yr}}$ & ${\sim 1.7 \times 10^{10}\,{\rm yr}}$                \\
\\
Model A1Ab & $2 \times 10^4 \Mo$ & $0.25\,{\rm pc}$ & 10 & $18.8\,{\rm kms^{-1}}$ & $8.2\,{\rm kms^{-1}}$ & $0.06{\rm pc}$ & ${1.7 \times 10^8\,{\rm yr}}$ & ${1.0 \times 10^6\,{\rm yr}}$ & ${\sim 1.7 \times 10^{10}\,{\rm yr}}$                \\                      
\\
Model A2Aa & $ 2 \times 10^4 \Mo$ & $0.5\,{\rm pc}$ & 5 & $13.1\,{\rm kms^{-1}}$ & $8.0\,{\rm kms^{-1}}$ & $0.18{\rm pc}$ & ${7.4 \times 10^8\,{\rm yr}}$ & ${1.9 \times 10^6\,{\rm yr}}$ & ${\sim 7.4 \times 10^{10}\,{\rm yr}}$             \\ 
\\
Model A2Ab & $2 \times 10^4 \Mo$ & $0.5\,{\rm pc}$ & 10 &$13.1\,{\rm kms^{-1}}$ & $5.7\,{\rm kms^{-1}}$ & $0.13{\rm pc}$ & ${7.4 \times 10^8\,{\rm yr}}$ & ${1.1 \times 10^6\,{\rm yr}}$ & ${\sim 7.4 \times 10^{10}\,{\rm yr}}$  \\ \hline
\\
\end{tabular}
\caption{Computed values of characteristic model parameters for a NSC with $2 \times 10^4$ stars with total mass $M_{\rm NSC}$; radius $R_{\rm NSC}$; cluster velocity $V_{\rm NSC}$; velocity of the BH $v_{\rm bh}$; the wandering distance for the BH from the NSC center $\Delta_{\rm bh}$; the two-body relaxation timescale calculated at the NSC half-mass radius $t_{\rm rh}$; the vector resonant relaxation timescale $t_{\nu {\rm RR}}$ and the estimated NSC evaporation timescale. Note that the vector resonant relaxation timescale is significantly shorter (by about two orders of magnitude) than the two-body relaxation timescale and the typical evaporation timescale is longer than the age of the universe for a range of model parameters.}
\end{table*}


 \section{Conclusions and Discussion}
\label{sec:conclusion}

In this work, we present a new channel for the formation of IMBHs in dense, gas-rich NSCs. NSCs, we show, can incubate the growth of an initially stellar mass black hole remnant into an IMBH. And this process can occur in NSCs throughout cosmic time. This continual BH formation mechanism can proceed and result in a range of final BH masses: if the process truncates pre-maturely, then growth stalls resulting in BHs with masses in the range $25 - 100 \Mo$. However, if the process continues unimpeded, then the mass of the BH, reaches between $100 - 1000 \Mo$, at which point it settles at the center after which subsequent growth is entirely limited by gas supply, comfortably yielding final IMBH masses of $10^{4-5} \Mo$. 

To summarize the model, the setting explored is here that of an initial $5 - 10 \Mo$ seed BH, embedded in a gas-rich NSC wherein it starts off in fluctuation-dissipation equilibrium with the cluster stars and gas. Following the dynamics of this seed BH as a test-particle, we find that it oscillates and wanders about the center due to interactions with cluster stars. Our proposed scenario is driven then by the supra-exponential wind-fed accretion of gas with finite optical depth onto this wandering BH facilitated by the dynamics of the BH in the host cluster. In the early growth phase, the random walking BH scatters off the stars which keeps the angular momentum in the accretion flow low. As is well known, any stronger-than-linear dependence of the accretion rate on the BH mass, leads to divergent growth of the BH mass in a finite time. This is the case in this early growth stage, and the timescale for divergence can be explicitly computed for the case of Bondi accretion in this setting. The physical conditions and constraints required for BH growth to proceed unimpeded in the NSC core region require that the cluster be and remain dynamically stable throughout. We find that the survival of the cluster is not a limiting factor on the time available for BH growth in the NSC environment. The shortness of the mass divergence timescale compared to other relevant dynamical timescales for the NSC permits this mass amplification process to proceed.

In addition to the availability of gas which is a fundamental constraint on accretion, circumventing the angular momentum barrier is also needed for mass to accrete onto the compact object. We find that the accretion in the NSC environment is not throttled by an angular momentum barrier by implicating two 
processes: the wandering of the BH due to scattering and vector resonant relaxation. As growth commences, random 2-body scattering of the BH by a small number of stars that are embedded in the core suffices to prevent the formation of an angular momentum barrier in the form an accretion disk. At this juncture, 
the disk-suppression mechanism is analogous to the concept of chaotic accretion introduced by \citet{king+06} in the context of MBH growth and spin evolution. Chaotic accretion refers to the situation where the direction of gas flow randomly changes with respect to the MBH. Here, it is the BH orbit that changes randomly with respect to the flow. And subsequent to this, we invoke rapid angular momentum relaxation of the orbit by vector resonant relaxation which operates in nearly spherical stellar systems like the NSC center where the BH orbits during its excursions. This well established process of vector resonant relaxation helps suppress the growth of an accretion disk around the wandering BH. Resonant relaxation, we argue, ensures suppression and circumvention of the angular momentum barrier. This occurs when the acceleration of the BH in the NSC falls below its equipartition value early on due to the accretion of pressure-supported (non-rotating) gas in the cluster, suppressing the build-up up of angular momentum in the flow and even possibly reversing it. Angular momentum relaxation occurs on the very short vector resonant relaxation time-scale (computed and shown for typical NSC models in Table~1), shorter than the mass divergence timescale, permitting circumvention of the angular momentum barrier for accretion. 

The initial growth phase is the critical stage for our picture, and it is also the regime where the stellar dynamics in the NSC can be modeled reliably analytically. Early growth therefore occurs via supra-exponential accretion in two stages: when the motion of the seed BH can be modeled as a simple harmonic oscillator as it in equipartition with the rest of the NSC and when the BH mass has grown significantly so that two body relaxation processes can no longer maintain equipartition and the motion is now that of a damped harmonic oscillator with progressively smaller excursions from the center before it finally gets massive enough that it can no longer be modeled as a test-particle and settles down into the center. After this final configuration is reached, continued growth by accretion is determined by the available gas supply. Depletion of the entire available gas reservoir can lead to a final IMBH mass of $\sim 10^{4-5} \Mo$.  

NSCs as sites for the formation of a wide mass range of black holes ranging from $25 \Mo$ to IMBHs over cosmic time proposed here can account for several new observational detections. These include LIGO-VIRGO detections \cite{LIGO+2020}, wherein close to half of the merging candidates have masses in excess of 25 $\Mo$. Premature termination of the growth channel in NSCs in Stage 2 could lead to the formation of such a population of BHs, even sources like GW190521 inferred to be the merger of two black holes with masses $85^{+21}_{14}\,\Mo$ and $66^{+17}_{18}\,\Mo$ with a total merged mass of $\sim 142\Mo$  \citep{LIGO+2020} in the so-called mass gap for direct stellar collapse (\cite{Heger+2002,Woosley2017}). \footnote{A complete list of secure LIGO-VIRGO merger events at {\it https://www.gw-openscience.org/eventapi/html/GWTC-1-confident/}}. The picture presented here also accounts naturally for the population of central and off-center IMBHs in low-mass galaxies that are being uncovered at present \cite{Mezcua+2019,Reines+20}. Meanwhile, if this process operates as described, then it also has important implications for the overall census of BHs in the universe, their occupation fraction in galaxies and enhancement in merger event rates expected for the LISA mission. In future work, we plan to implement and add this mechanism into our current semi-empirical black hole growth framework \citep{Ricarte&Natarajan2018b} that tracks the formation and fueling of black hole seeds over cosmic time. 

We have demonstrated that NSCs can incubate the formation of IMBHs throughout cosmic time tracing the mass assembly process using reliable analytic models.  There are several caveats and limits to the modeling approach adopted here. We have made a couple of approximations whose validity while adequately justified, needs to be examined in further detail preferably using tailored numerical simulations. For instance, carefully computing the transition from Stage 1 (random walk executed about the NSC center as harmonic motion) to Stage 2 (when motion is that of a damped harmonic oscillator) of the process when equipartition can no longer be maintained would be useful to explore further numerically. In this treatment, we have also not taken into account the larger scale galactic environment in which the NSC is embedded, and we have treated the NSC as a closed box cocoon. Including phenomena on much larger physical scales and exploring the cross-talk between the environment and the NSC is the beyond the scope of our current study. Fully self-consistent simulations are not currently possible, as they are computationally prohibitive given the coupling of physical scales, astrophysical processes and large-scale environmental effects that need to be taken into account. In future work, we plan to explore the use of more sophisticated models and couple them to idealized simulations to further study this mechanism in detail.

\section*{acknowledgements} 

I gratefully acknowledge many useful conversations with the late Tal Alexander on this problem. My own thinking on accretion physics was greatly sharpened by many discussions and work over the years with Phil Armitage, Marta Volonteri, Chris Reynolds, Jim Pringle and Mitch Begelman - to them, I am deeply thankful. 

\section*{Data Availability Statement}

All the data generated as part of this project and calculation are provided in the paper. 

\bibliography{smbh_master}

\end{document}